\documentstyle[psfig]{mn}				
\newif\ifAMStwofonts

\newcommand{\mc}{\multicolumn}
\newcommand{\U}{U$_{300}$}
\newcommand{\B}{B$_{450}$}
\newcommand{\V}{V$_{606}$}
\newcommand{\I}{I$_{814}$}
\newcommand{\omg}{$\Omega_0$}
\newcommand{\tmax}{$t_{max}$}

\def\etal{{\it et al.~\/}}

\def\ltsima{$\; \buildrel < \over \sim \;$}
\def\simlt{\lower.5ex\hbox{\ltsima}}
\def\gtsima{$\; \buildrel > \over \sim \;$}
\def\simgt{\lower.5ex\hbox{\gtsima}}

\title{The Hubble Deep Field Reveals a Supernova at z$\simeq$0.95}

\author[Mannucci \& Ferrara]
	{F.~Mannucci,$^1$ 
	 A. Ferrara,$^2$\thanks{Affiliated to Joint Institute for
	Laboratory Astrophysics, University of Colorado,
		   Boulder, CO 80309, USA}\\
    $^1$C.A.I.S.M.I.--C.N.R., Largo E.Fermi 5, 50125 Firenze,  Italy\\
	$^2$Osservatorio Astrofisico di Arcetri, Largo E.Fermi 5, 50125 Firenze,  Italy
}
\date{Accepted . Received }
\pagerange{\pageref{firstpage}--\pageref{lastpage}}
\pubyear{1999}

\begin{document}

\label{firstpage}

\maketitle

\begin{abstract}
We report the discovery of a variable object in the Hubble Deep
Field North (HDF-N) which has brightened,
during the 8.5 days sampled by the data, by more than 0.9 mag in \I\ 
and about 0.7 mag in \V, remaining stable in \B. Subsequent observations of the HDF-N
show that two years later this object has dimmed back to about its original 
brightness in \I.
The colors of this object, its brightness, its time behavior in the various 
filters and the evolution of its morphology are consistent with being 
a Type Ib supernova in a faint galaxy at $z\sim $0.95.
\end{abstract}


\section{Introduction}
The discovery of supernovae (SNe) in the early universe is of great
interest because they can provide a wealth of information about 
cosmological parameters and the cosmic star formation history. 
It is now believed that the star formation activity in the universe 
very likely started in small objects  which later on merged to 
form larger units (Couchman \& Rees 1986, Ciardi \& Ferrara 1997, 
Haiman \etal 1997, Tegmark \etal 1997, Ferrara 1998). 
Unless the IMF at high-$z$ is drastically different from the
local one, some of the formed stars will end their lives as
SNe. Detecting high-$z$ SNe would be of primary importance to clarify how     
reionization and rehating of the universe proceeded \cite{ciardi97}, 
and, in general, to derive the star formation history of the universe 
(Sadat \etal 1998, Madau, Della Valle \& Panagia 1998)
and pose constraints on the IMF and chemical enrichment of the universe. 

Great effort has been spent in this search \cite{garnavich98,perlmutter98}
and many SNe have been found up to a redshift of $z=1.20$ \cite{aldering98}
when the universe had only about half of its present age.
The HDF images \cite{williams96} are the among the deepest taken to date and 
in principle could contain SNe up to $z\sim$3.
Two SNe\footnote{These two SNe are different objects with respect to the
ones reported in this work} were actually detected by Gilliland \& Phillips 
(1998) and Gilliland, Nugent \& Phillips (1999)
by comparing the primary HDF-N data with second-epoch images taken 
two years after.
The primary observations of this field were taken using four optical filters 
centered at 3000\AA\ (\U), 4500\AA\ (\B), 6060\AA\ (\V) and 8140\AA\ (\I) 
during a time span of about 10 days in December 1995.
Although the distribution of the data along this period is 
not uniform, the overall time coverage is good enough to detect
objects with significant variations on time scales of a few days.
High redshift SNe cannot be identified near their maximum in such a short time
span because they vary too slowly, but soon after the explosion they evolve 
fast enough to be detected. 

How many SNe can be expected in the HDF-N? 
This number can be estimated using the 
computations by Marri \& Ferrara (1998) and Marri \etal (1998)
for future Next Generation Space Telescope (NGST) surveys. 
Scaling their results for a flat CDM+$\Lambda$ with $\Omega_M=0.4$
to the HDF-N area 
($\sim$ 5 sq. arcmin) and assuming a surveying time of $\sim 8$ days
(see below), we obtain an expected number $\sim 0.34$ SNe from massive 
stars (SN types Ib/II). As a comparison,
scaling the analytical estimates in Miralda-Escud\'e \& Rees (1997) to the
HDF-N, one obtains a similar expected rate of 0.17 SNe in the HDF-N.
These values are large enough to encourage a new analysis of the HDF-N. 

\section{Object selection, photometry and morphology}

The original observations of the HDF-N in each filter consisted of about 300
images taken in several (from 9 to 11, depending on the filter)
positions on the sky (dither positions).
We have divided these images
into a few subsequent groups, three for \V\ and \I\ and two for \B.
The \U\ band, which is intrinsically less efficient, was not considered.
The HDF-N was observed again two years later in \U\ and \I\, 
and the latter image (I4 in Table 1)
is deep enough to be used for this project.  

Where possible, images taken in the same position on the sky were not split
into different groups so that we could make effective use of those 
combined
by the HDF working team \cite{williams96} for each dither position, 
and made available on the web. 
Some dither positions in \V\ and \I\ 
contain images too widely separated in time and therefore have been split. 
In this case the frames were reduced by the automatic pipeline provided by 
the ST-ECF web site. In all cases warm and bad pixels were rejected 
using the standard routines.

Given the small number of images, between 3 and 6, present in some of
the final groups, it was not possible to use the {\em drizzle} algorithm
\cite{fruchter97} often used for the WFPC2 data reduction. We have
chosen to
resample all the images to a pixel of 0.05 arcsec (half of the original
WF pixel scale) and use the IRAF task IMCOMBINE to make the final
combinations while rejecting deviant pixels.
This procedure is more efficient than drizzle
in rejecting any residual warm and bad pixel; this is especially true
for the final \V\ and \I\ images which contain 5 dither images each.
Table 1 lists the properties of the resulting images with their 
time coverage and limit magnitude.
\begin{table*} 
\begin{minipage}{10cm}
\caption{ Grouping of the images and properties of 2-584.2}
\begin{tabular}{cclcccrc}
    Filter & 
	Image & 
	Dither pos.\footnote{Dither positions as in Williams \etal 1996.
	When present, A and B refer to the first and second part of the 
	split positions}&
	Exp. time & 
	Time range & 
	Limit\footnote{3$\sigma$ limit for point sources}& 
	\mc{2}{c}{2-584.2}  \\
     &       
	 &  
	 &  
	 (hours)  & 
	 (days)   & 
	 & 
	 Mag & 
	 SNR\footnote{Signal-to-noise ratio of the photometry}         \\
      &      &                  &    &             &       &         &     \\
F450W &Total &                  & 33.5 & 0.00--9.21  & 29.72 &   28.73 & 5.1 \\
      &B1    & 6A+9+4+5+3+8+1+2 & 28.2 & 0.00--3.15  & 29.64 &   28.75 & 7.5 \\
      &B2    & 7+6B             &  5.3 & 8.34--9.21  & 28.72 &   28.41 & 3.8 \\
      &      &                  &      &             &       &         &     \\
F606W &Total &                  & 30.3 & 1.48--10.07 & 30.07 &   28.78 & 7.2 \\
      &V1    & 3+2+6A+4A        &  5.6 & 1.48--5.56  & 29.16 &   29.14 & 3.1 \\
      &V2    & 9A+5A+7+1+11+10  & 14.7 & 6.43--8.51  & 29.68 &   28.93 & 6.0 \\
      &V3    & 8+6B+4B+9B+5B    & 10.0 & 8.57--10.07 & 29.47 &   28.53 & 7.2 \\
      &      &                  &      &             &       &         &     \\
F814W &Total &                  & 34.3 & 3.21--10.10 & 29.46 &   28.63 & 4.8 \\ 
      &I1    & 2+1+6            & 10.0 & 3.21--4.35  & 28.80 &$>$28.80 & 2.8 \\
      &I2    & 4A+9+3+5A        & 14.5 & 4.69--6.30  & 29.00 &   28.76 & 3.8 \\
      &I3    & 8+7+11+4B+5B     &  9.8 & 6.63--10.10 & 28.79 &   28.05 & 6.0 \\
	  &I4    &                  & 15.6 & 735--738    & 29.00 &$>$29.00 & 2.9 \\
\end{tabular}
\end{minipage}
\end{table*}

Two tests were performed to check the data reduction results:
i) the flux of a few compact sources from the list in M\'endez \etal (1996)
was measured in each of the resulting images to verify the
constancy of the photometry;  
ii) as discussed, for example, in Ferguson (1998), the effective limiting
magnitude depends on the object size as much as on its total magnitude.
The detection limits for point sources in the images were 
measured by adding stars and detecting them using FOCAS \cite{valdes82}: 
the derived values for the 80\% completeness
level are in good agreement (within
0.1 mag) with the values in Williams \etal (1996) once their 10$\sigma$ limits 
in an aperture of 0.5 arcsec are scaled for the exposure time and
to about 3$\sigma$ in 0.30 arcsec, an aperture yielding
the highest SNR for point source photometry (eg., Thompson 1995).
These values are listed in Table 1; all magnitudes are in the AB system.

The three combined \V\ images (V1, V2 and V3), showing the best sensitivity
(3$\sigma$ limits between 29.1 and 29.7) and time coverage (about 8.5 days),
were examined for variable objects having a monotonic trend, either 
brightening or dimming. A few interesting objects were selected;
the most remarkable one is in chip 2 of the WF camera and is 
present in the Williams \etal (1996) catalog with the entry number 584.2. 
The J2000 coordinates of this object (2-584.2) are 12:36:49.357 +62:14:37.50.
In the total images, this object has flat \B, \V\
and \I\ colors and is undetected in \U; however it cannot be
classified as a \U\ drop-out being too faint to show a strong enough
break between \U\ and \B.

Table 1 and Fig. 1 show the time evolution of the photometry of 2-584.2.
Its magnitudes in the various images were measured inside a circular aperture 
of 0.3 arcsec, corresponding to about twice the PSF FWHM, and corrected
to an infinite aperture. This object shows a strong brightening both in \V\
and \I, while the photometry in the \B\ band is consistent with a 
constant value. The errors shown in Fig. 2 are derived under the
assumption of Poisson noise inside the photometric aperture.
\begin{figure}
\centerline{\psfig{figure=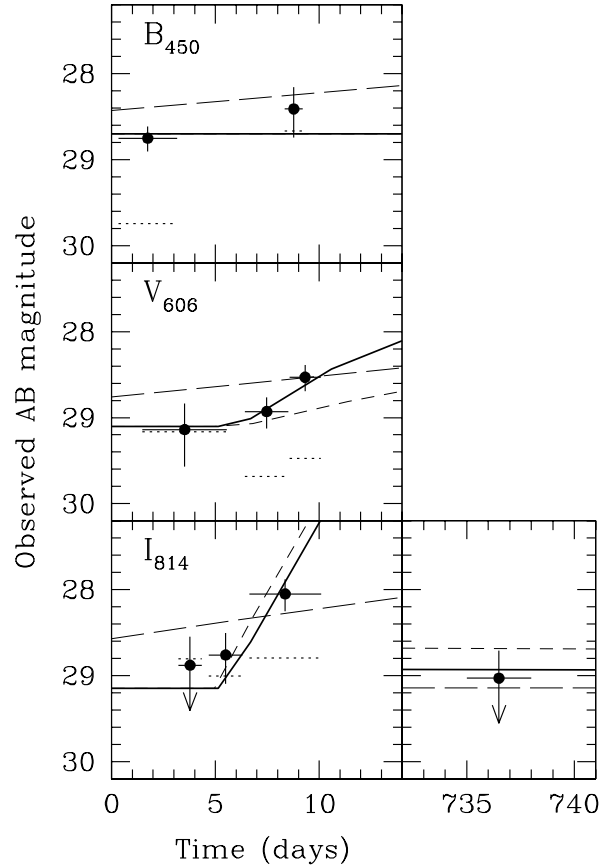,width=8cm}}
\caption{ 
Photometry of 2-584.2 in \B, \V, and \I\ as a function of 
the (observer) time; day=0 corresponds to the first observation.
Each point shows the 1 $\sigma$ photometric errors and the time span 
covered by the data; the dotted lines are the 3 $\sigma$ limits
of each image. The the arrows in 
the \I\ panels show the 3 $\sigma$ upper limits.
The solid thick lines are a fit obtained from a K-corrected, time-dilated
SNIb light curve at $z= 0.95$ with \tmax=34.0, E(B-V)=0.04 and \omg=0.1.
The spectrum and the time evolution of a SNIb simultaneously fit the
magnitudes of the variable object and its evolution in the three filters.
The underlying galaxy results to have \B=28.70, \V=29.10 and \I=29.15.
Also shown are the curves for a SNIa at $z\simeq 1.3$ (short dashed) 
and a SNII at $z\simeq 1.1$ (long dashed) which do not provide an
acceptable match.
}
\end{figure}

We have computed the statistical significance of the detected variation. 
For the \V\ band we consider the two differences V2--V1 and V3--V2; 
in the \I\ band, as the object shows no significant variation among 
I1, I2 and I4, we coadd these three images and compare the result with I3.
By comparing the differences in the photometry with the quadratic sum of
the errors, we find a joint probability for these three variations 
to arise from noise
of 1.2$\times 10^{-5}$. In the HDF-N there are about 1200 objects between 
\I=27 and \I=29 \cite{williams96}; 
therefore we expect one such spurious event only every 70 Hubble Deep Fields.

Fig. 2 shows the images of this object in the various bands
\footnote{ Images of 2-584.2 can also be found at 
http://www.arcetri.astro.it/$\sim$filippo/sn/sn.html}
and its brightening in \V\ and \I.
In the total images the object is marginally resolved in
\B\ and \V, whereas in \I\ it is consistent with a point source. 
The images also give the impression of a sharpening of 2-584.2 with
time: the object seems more extended at the beginning of the observations
than at the end, as if a bright core were emerging in \V\ and \I.
Its faintness prevents us from studying its morphology in detail;
nevertheless, its extension can be roughly measured by fitting circular 
gaussians to each image. The results of this procedure supports the
visual impression of a sharpening
of the object with time: the FWHM of the best-fitting gaussian goes 
from 0\farcs 33$\pm$0\farcs 08 to 0\farcs 18$\pm$0\farcs 02 in \V\ and 
from 0\farcs 40$\pm$0\farcs 10 to 0\farcs 19$\pm$0\farcs 02 in \I\, 
while the value expected for point sources is about 0\farcs 15.
\begin{figure*}
\centerline{\psfig{figure=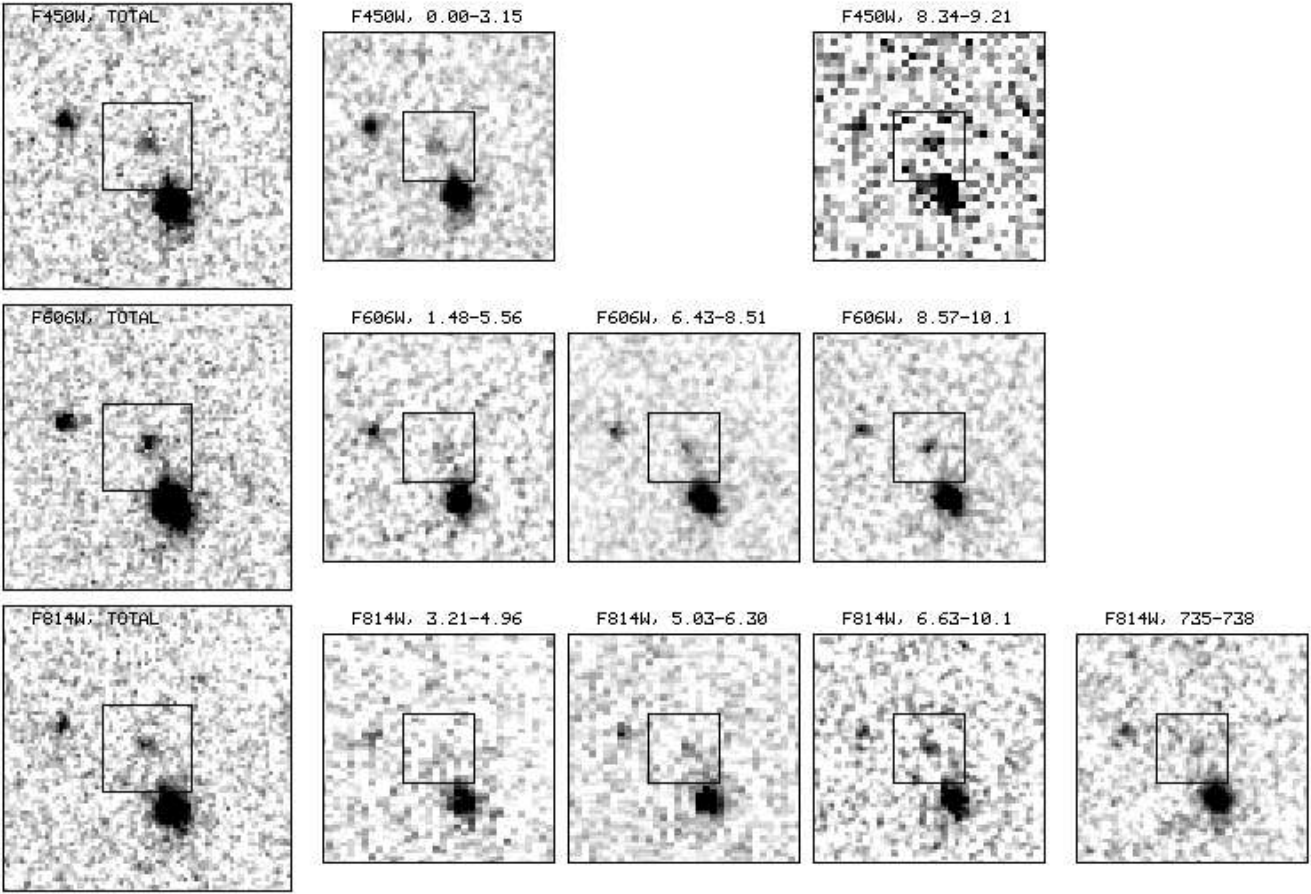,width=17.7cm}}
\caption{ 
Images of 2-584.2 in the \B, \V\ and \I\ bands.
The three images on the left are the total HDF-N images by
Williams \etal 1996 with a pixel scale of 0.04 arcsec/pixel. 
On the right the time sequence in each filter with pixel scale 
of 0.05 arcsec/pixel showing the brightening of the object in \V\ and \I\
and its following dimming. 
Each image is labelled by the filter and the time range of the 
observations in days.
The size of the central squares is 1 arcsec. 
}
\end{figure*}

\section {The light curve}

Among the objects that could show variability at this high galactic
latitude and faint flux level, AGNs and SNe are the most plausible candidates.
The shape of detected time variation and the possible development of a 
bright core in an underlying galaxy strongly suggest the identification of
2-584.2 with a SN observed soon after its explosion.

The expected evolution of the apparent magnitude of a SN
can be derived from template light curves, spectra and absolute magnitude 
(as discussed below) once its distance is known and  K-corrections (due to
the narrowing of the filter passband in the restframe of the source
and redshifting of the emitted photons from the source to the
observer) are computed.     
For high redshift objects  
the time in the observer frame, $t_{obs}$, is related to the time in the
supernova rest frame by $t_{obs}=(1+z)t_{rest}$; this produces a 
time dilation of the high redshift SN light curve. 

Once the SN template is given, four parameters are left 
to simultaneously fit the data in the various filters:
the SN redshift, $z$, the time of the maximum light, \tmax, 
the color excess due to dust, E(B-V),
and the cosmological matter density parameter, \omg\ (we assume $\Lambda=0$). 
Some uncertainty remains associated with this fit as 
pre-maximum light curves and spectra (particularly in the UV) 
are generally not very well determined 
because they are available only for a handful of objects.
The SN flux must then be added to the host galaxy flux in each
filter. The observed time variation of 2-584.2 shows that 
the SN becomes quickly dominant in \V\ and \I\, while the galaxy
produces most of the \B\ flux. 

We now consider in turn the possibility that the detected variable
source is a Type Ia (SNIa), a Type II (SNII) or a Type Ib (SNIb) supernova. 
SNIa have been shown to be reliable standard candles, as their intrinsic
luminosities can be accurately determined \cite{phillips93}. 
Pre-maximum optical (CTIO) and UV (IUE) spectra of SN1990N 
\cite{leibundgut91}, 
show that the flux drops off sharply below $2600$\AA; therefore,
as soon as $z\simgt 0.7$, their \B\ flux in negligible 
as the cut off is redshifted into this band,
while the mere detection of variation in the \V\ band puts the upper limit 
$z \simlt 1.6$. 
The luminosity of a SNIa evolves in time according to a 
light curve \cite{leibundgut88,doggett85} which shows a fast rise to 
the maximum (3.6 mag in 15 days), with slight differences between 
the (rest frame) B- and V-band. 
When fitting the data for 2-584.2 with the time-dilated, K-corrected 
light curves, we find impossible to simultaneously match the
\V\ and \I\ data points
(in this case the \B\ data only give information on the galaxy magnitude), 
as shown in Fig. 1.
Since the object is relatively faint, 
this implies high SNIa redshifts ($z\simgt 1.3$) 
for which colors cannot be reproduced by using appropriate K-corrections. 
We conclude that identification of 2-584.2
with a SNIa can be safely ruled out.  

We repeat the same procedure for SNII. These objects cannot be used 
as standard candles as their peak abolute luminosities are known to cover 
a wide range, from $M_B\sim -14$ to $M_B\sim -19$ \cite{patat94}.
SNII are usually divided into two classes \cite{doggett85}, 
one (SNII-P) showing a slow
pre-maximum brightening and a plateau in the after-maximum decline, the
other (SNII-L) a faster brightening and a linear decline.
In both cases the pre-maximum spectra can be approximated
by a black-body with a temperature $T_{BB}\sim 25000$~K \cite{kirshner90}
without any UV cut-off. 
By comparing the expected light curves with the data (see Fig. 1) 
we can safely exclude 
both classes because (i) their brightening is too slow 
and (ii) their blue spectrum makes them too luminous in \B.

SNIb yield a good agreement with the data. These objects 
closely resemble SNIa in terms of time evolution and 
spectra but are typically 1-2 mag fainter \cite{wheeler85,kirshner90}
and show a larger spread in the maximum brightness.
While SNIa are found in all types of galaxies and derive from old stars, 
the SNIb are only detected near regions of active star formation and 
their progenitors should be young massive stars. 
As shown in Figure 1, both the time evolution of 
2-584.2 increasing from \B\ to \V\ to
\I\ and its apparent magnitudes are
easily reproduced by SNIb light curves \cite{kirshner90}: 
acceptable fits can be obtained for $0.90 < z < 1.02 $
which makes this object one of the most distant SN
observed to date, 
while the best agreement is found for $z=0.95$, 
\tmax=$34.0\pm1$~days ($\sim 12$ rest-frame days after the end of 
the observations), low \omg\ values and 
moderate reddening $0\le$E(B-V)$\le0.05$ \cite{seaton79}. 
This values of the reddening,
corresponding up to about 0.29 mag of extinction in \V\ and 0.22 in \I,
could also be accounted for by a SNIb fainter than the average
and without extinction.
Lower values for the fitting redshift tend to select low values of
\omg\ and low extinction, viceversa for higher redshifts.
The galaxy, showing flat \B-\V\ and \V-\I\ colors and luminosites indicating 
a SFR of about 0.2 M$_{\odot}$/yr \cite{madau98}, is consistent 
with a star forming dwarf. About 20\% of the flux in the \I\ image
taken two year later (I4) is still due to the SN.

We therefore conclude that: (i) SNIa are too bright and red and
SNII are too slow and blue to be viable candidates for 2-584.2;
(ii) a SNIb naturally reproduces the time evolution, 
brightness and colors of the variable object. 
This type of study, when applied to high sensitivity, long time span
future observations, could produce
a significant sample of high-$z$ SNe that can be  
succesfully used to constrain the star formation
history and the geometry of our universe. 

\section*{Acknowledgments}

We are indebted to P. Hoeflich and B. Leibundgut for providing
supernova spectra and light curves, and to R. McCray, M. Salvati, L.
Pozzetti and N. Panagia for useful discussions.  
We also thank the referee, R. Ellis, for insightful comments.
We thank  STScI/ST-ECF for the implementation of their data
archive and for support during this work.
FM acknowledges a partial support from ASI grant ARS-96-66;
AF acknowledges partial support as a Visiting Fellow at JILA.\\

\bsp

\label{lastpage}
 
\end{document}